\newcommand{\km}{\hbox{ km}}
\newcommand{\s}{\hbox{ s}}
\newcommand{\mpc}{\hbox{ Mpc}}
\newcommand{\gyr}{\hbox{ Gyr}}

\newcommand{\cfrac}[2]{\textstyle{\frac{#1}{#2}}}

\documentclass[aps,prl,twocolumn,groupedaddress,showpacs,nofootinbib,
nobibnotes,twoside]{revtex4}
\usepackage{graphicx} 
\usepackage{hyperref}
\usepackage{latexsym}
\preprint{FERMILAB--PUB--04/nnn--T}
\catcode`\@=11
\def\ps@fnal{\def\@oddhead{\textsf{FERMILAB--Pub--04/368--T \hfil \thepage}}
\def\@evenhead{\thepage \hfil \textsf{FERMILAB--Pub--04/368--T}}}
\relax

\begin{document}

\pagestyle{fnal} 
%


\title{Undulant Universe} 


\author{Gabriela Barenboim}
\email[]{Gabriela.Barenboim@uv.es}
\affiliation{Departament de F\'{\i}sica Te\`{o}rica,
 Universitat de Val\`{e}ncia, 
Carrer Dr.~Moliner 50, E-46100 Burjassot (Val\`{e}ncia),  ÊÊSpainÊ}

\author{Olga Mena Requejo}
\email[]{omena@fnal.gov}
\affiliation{Theoretical Physics Department, Fermi National 
Accelerator Laboratory,\\ P.O.\ Box 500, Batavia, Illinois 60510 USA}

\author{Chris Quigg}
\email[]{quigg@fnal.gov}
\affiliation{Theoretical Physics Department, Fermi National 
Accelerator Laboratory,\\ P.O.\ Box 500, Batavia, Illinois 60510 USA}


\date{\today}

\begin{abstract}
If the equation of state for ``dark energy'' varies periodically, the
expansion of the Universe may have undergone alternating eras of
acceleration and deceleration.  We examine a specific form that
survives existing observational tests, does not single out the present 
state of the Universe as exceptional, and suggests a future
much like the matter-dominated past: a smooth expansion without a final
inflationary epoch.
\end{abstract}

\pacs{98.80.Cq, 95.35.+d, 98.70.Vc \hfill 
\fbox{\textsf{FERMILAB--PUB--04--368--T}}} 

\maketitle

The discovery in measurements of distant supernova redshifts that the
Universe is expanding at an accelerating
pace~\cite{Riess:1998cb,Perlmutter:1998np} has been reinforced and
extended by detailed observations of anisotropies in the cosmic
microwave background~\cite{Bennett:2003bz} and broad surveys of
large-scale structure~\cite{aussies,sloan}.  The weight of
observational evidence points to a flat Universe whose mass-energy 
includes
5\% ordinary matter and 22\% nonbaryonic dark matter, but is dominated by the
``dark energy'' identified as the motor for accelerated
expansion.\footnote{For three complementary views of today's
concordance cosmology, see
Refs.~\cite{Peebles:2002gy,Freedman:2003ys,Trodden:2004st}.} The
inferred Universe has been variously described as
extravagant~\cite{kirshner} and
preposterous~\cite{Carroll:2001xs}---extravagant, because we are
acquainted with so little of the stuff that makes up the Universe, and
preposterous because today's rough balance between matter and energy
was unforeseen and might be a once in a lifetime occurrence, not just 
for observers, but also for  the Universe!

Because the fossil record is spotty, and we are still learning how best
to read it, there is much room for interpretation~\cite{Bridle:2003yz}.
The most economical description of the cosmological measurements
attributes the dark energy to a cosmological constant in Einstein's 
equation---an omnipresent and invariable vacuum energy
density~\cite{Krauss:1995yb}.  A dynamical option is to suppose 
that a  cosmic scalar field, called quintessence, changing 
with time and varying across space, is slowly approaching its ground 
state~\cite{Caldwell:1997ii}.

On the cosmological constant interpretation, the vacuum energy density would have made a
scant contribution to the energy portfolio of the young small Universe,
but the future Universe would grow so quickly as to be essentially
empty of matter.  Quintessence models admit many destinies. 
Neither proposal naturally explains why matter and
vacuum energy should be of comparable importance at this moment in
cosmic history.  Responses to the ``why now?''\ question range from
anthropic rationalizations~\cite{Susskind:2003kw} 
to cyclic cosmologies~\cite{Steinhardt:2002ih}.

In this Letter, we explore the possibility that the equation of state 
of the vacuum energy  is an oscillatory function of the scale of the 
Universe. We shall show that a simple \textit{Ansatz} compatible with 
 existing observations responds naturally to the ``why now?''\ problem and 
 connotes a cosmic destiny similar to that of a matter-dominated 
 critical-density universe. In common with a cosmological constant, but in
 distinction to quintessence (which embodies a frozen field), the dynamical 
 origin of such a universe could 
have been present at early times.
 

The expansion of the universe is determined by the Friedmann equation,
\begin{equation}
    H^{2} \equiv ({\dot{R}}/{R})^{2} = {8\pi 
    G_{\mathrm{N}}\rho}/{3} - {k}/{R^{2}} + {\Lambda}/{3}\;,
    \label{eq:friedmann}
\end{equation}
where $H$ is the Hubble parameter, $R$ is the cosmological  
scale factor, $G_{\mathrm{N}}$ is Newton's constant, $\rho$ is the 
energy density, $k = 
(+1,0,-1)$ is the curvature constant,\footnote{If $\Lambda = 0$, the 
curvature constant determines destiny. For $k=+1$ (closed Universe), 
the Universe recollapses in finite time; for $k=0$ (flat) and $k=-1$ 
(open), the Universe expands without limit.} and $\Lambda$ is the 
cosmological constant. It is convenient to define the dimensionless 
scale factor, $a = R/R_{0}$, where the subscript $0$ denotes the 
value at the current epoch. The critical density, defined from 
(\ref{eq:friedmann}), is $\rho_{c} = {3H^{2}}/{8\pi G_{\mathrm{N}}}$.
The dimensionless cosmological density parameter is defined relative to the 
critical density as $\Omega_{\mathrm{tot}} = \rho/\rho_{c}$ at any 
epoch. 
We express the rate of change of the Hubble parameter through the 
deceleration parameter,
\begin{equation}
    q \equiv - \frac{1}{H^{2}} \, \frac{\ddot{R}}{R} = 
    \frac{\Lambda}{3H^{2}} - \frac{4\pi 
    G_{\mathrm{N}}}{3H^{2}}\,(\rho + 3p)\;,
    \label{eq:decel}
\end{equation}
where $p$ is the isotropic pressure. If we define 
$\Lambda = 4\pi G_{\mathrm{N}}\rho_{\Lambda}$ and introduce the 
equation of state $w_{i} = p_{i}/\rho_{i}$ for any component of the 
universe, we can recast the deceleration parameter as
\begin{equation}
    q = \cfrac{1}{2}\sum_{i}\Omega_{i}(1 + 3 w_{i}) = \cfrac{1}{2} 
    \left(\Omega_{\mathrm{tot}} + 3 \sum_{i}\Omega_{i}w_{i} \right)\;.
    \label{eq:decel2}
\end{equation}
The equation of state of pressureless matter is $w_{m}=0$, and that of 
radiation is $w_{r}= \cfrac{1}{3}$. We see by inspection of 
Eq.~(\ref{eq:decel}) that $w_{\Lambda} = -1$.

The $\Lambda$CDM proposal is parsimonious in its introduction of a 
single parameter, $\Omega_{\Lambda}$, but offers no explanation for 
the peculiar circumstance that $\Omega_{\Lambda} \approx \Omega_{m}$ 
at the current epoch---and no other---in the history of the universe. 
It is interesting to probe the range of interpretations that reproduce 
the observed features of the universe.


We investigate here the possibility that the physical characteristics 
of the vacuum energy vary with time, specifically with the number of 
$e$-foldings of the scale factor, with an equation of state 
\begin{equation}
    w_{v}(a) = -\cos(\ln a)
    \label{eq:oureos}
\end{equation}
that matches the inference that $w_{v0} \approx -1$ in the
current universe.\footnote{Equations of state involving 
$\cos(\ln{a})$ have been explored, to a different end, in Ref.~\cite{Feng:2004ff}.} We assign the vacuum energy a
weight $\Omega_{v0}=0.7$, in line with observations, and take 
$\Omega_{m0}=0.3$ and $\Omega_{r0} = 4.63\times10^{-5}$. The present-day expansion rate is 
$H_{0} = 100\,h\km\s^{-1}\mpc^{-1}$, with $h = 
0.71^{+0.04}_{-0.03}$~\cite{Eidelman:2004wy}.  

\begin{figure}
    \includegraphics[width=9cm]{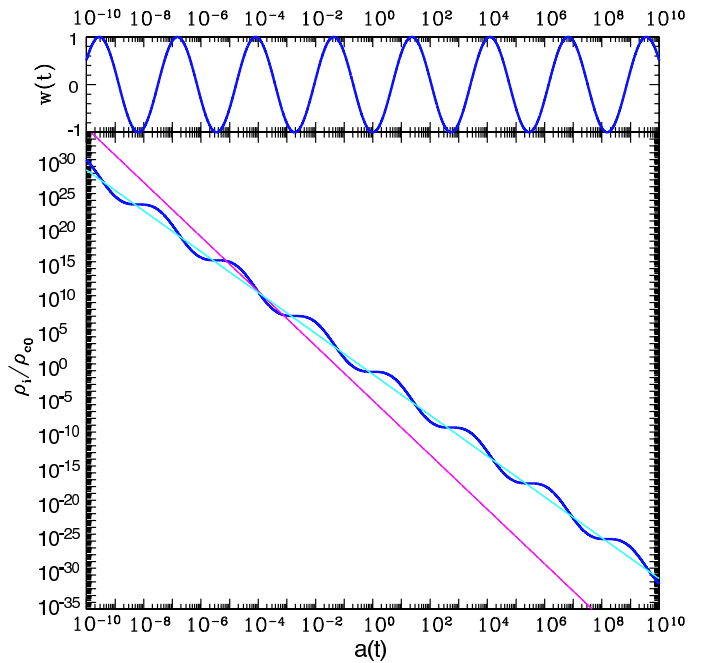}
\vspace*{-12pt}
\caption{Lower panel: Evolution of the matter (thin cyan), radiation 
(magenta, steepest line), and 
vacuum (thick blue) energy densities in the undulant universe, normalized to the critical density 
$\rho_{i}/\rho_{c0}$,  versus the scale factor $a(t)$. Upper panel: 
Equation of state, Eqn.~(\ref{eq:oureos}), of the undulant 
vacuum.}
\label{fig:fish}
\end{figure}

Because over one period the equation of state (\ref{eq:oureos})
averages to zero (the equation of state of pressureless matter), the
cosmic coincidence problem is resolved. We plot in Figure~\ref{fig:fish} 
the normalized energy densities of matter, radiation, and vacuum 
energy as functions of the scale parameter $a$. These are given in 
terms of the normalized densities now as
$\rho_{m}/\rho_{c0} = \Omega_{m0}/a^{3}$, $\rho_{r}/\rho_{c0} = 
\Omega_{r0}/a^{4}$, and $\rho_{v}/\rho_{c0} = g(a)\Omega_{v0}/a^{3}$, 
where 
\begin{equation}
    g(a) =  e^{3\int_a^1 d{a^\prime}\,w(a^\prime)/a^{\prime} }
  =   e^{3 \sin(\ln{a})}\;.
    \label{eq:gofa}
\end{equation}
Looking back in time to the epoch of big-bang nucleosynthesis at $a 
\approx 10^{-10}$, and forward to $a = 10^{+10}$, we see that the 
vacuum energy density crosses the matter density
every $\pi$ $e$-foldings of the scale factor. 
These regular crossings stand in sharp contrast to the $\Lambda$CDM 
cosmology, in which $\Lambda_{v} \approx \Lambda_{m}$ only in the 
current epoch. Periodically dominant dark energy   is 
in the spirit of  Refs.~\cite{Dodelson:2001fq,Griest:2002cu}.

The Hubble parameter  is now given by
\begin{equation}
    H(a) = H_0 \sqrt{ \frac{\Omega_m}{a^{3}} + \frac{g(a)
    \Omega_v}{a^{3}} +\frac{\Omega_r}{a^{4}} }\;,
    \label{eqn:hubble}
\end{equation}
and the current age of the universe, $t_{0}  = \int_0^1  da/H(a) 
a$, is $13.04\gyr$, to be compared with $13.46\gyr$ in the 
$\Lambda$CDM model. Both values are in good agreement with the age of 
$(12.9 \pm 2.9)\gyr$ inferred from  globular clusters~\cite{Cepa:2004bc}. 
By calculating the time to reach a given scale factor, we can 
determine the history and future of the universe.
During the radiation dominated era, which corresponds to $a 
\lesssim 10^{-5}$, $a(t) \propto t^{1/2}$; when matter dominates, $a(t) 
\propto t^{2/3}$. 

\begin{figure}
\includegraphics[width=8.75cm]{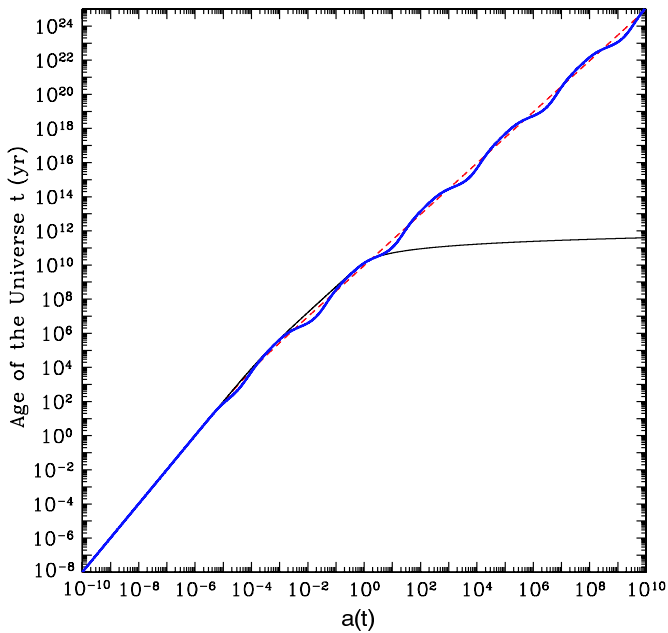}
\caption{Evolution of the scale factor $a(t)$ in three 
cosmologies: the canonical $\Lambda$CDM model (thin black line); a critical universe (SCDM
model) with $\Omega_{m} = 1$ (dashed red line); and the periodic equation of
state (\ref{eq:oureos}) (thick blue line).}
\label{fig:scalefactor}
\end{figure}

We show the results for three cosmologies in  
Figure~\ref{fig:scalefactor}. 
The dashed (red) line corresponds to the ``standard cold dark 
matter'' (SCDM) cosmology that was canonical before the discovery of 
the accelerating universe. The thin solid (black) line shows the 
$\Lambda$CDM cosmology, in which the present epoch marks the beginning of 
a final inflationary period that leads to an empty universe in which 
matter is a negligible component. The heavy (blue) line shows the 
prediction of Eqn.~(\ref{eq:oureos}). In the recent past, the 
periodic equation of state matches the behavior of the $\Lambda$CDM 
cosmology, but in the future it undulates about the SCDM prediction. 

\begin{figure}
\includegraphics[width=8.5cm]{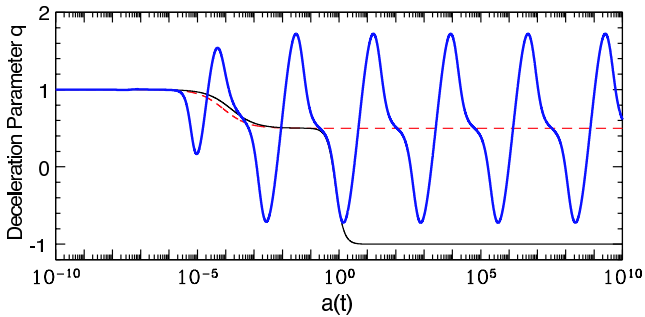}
\caption{The deceleration parameter as defined in Eq.~(\ref{eq:decel2}) 
for the undulant universe (thick blue line), $\Lambda$CDM model (thin 
black line) and SCDM 
model (dashed red line).}
\label{fig:decel}
\end{figure}

The expansion of the \textit{undulant universe} is characterized by
alternating periods of acceleration and deceleration shown by the
deceleration parameter in Figure~\ref{fig:decel}.  For scale factors
$a$ between $0.1$ and $1$, the periodic equation of state tracks the
behavior of the $\Lambda$CDM cosmology.  But whereas $\Lambda$CDM is
about to enter an era of sustained acceleration, the average behavior
of the undulant universe tracks that of SCDM.

We have performed a number of checks to verify that the periodic
equation of state is consistent with existing observations.  At BBN, we
find $\Omega_{v} \approx 2 \times 10^{-5}$, which respects the bound
$\Omega_{v} < 0.045$ ($2\sigma$ limit)~\cite{Bean:2002sm}.  The model
reproduces the luminosity distance modulus of the Supernova Search
Team's gold and silver samples~\cite{Riess:2004nr} and the x-ray gas
mass fractions determined by the Chandra X-ray
Observatory~\cite{chandra,Allen:2004cd,Rapetti:2004aa}.  We find no
significant difference between the linear growth
factors~\cite{Linder:2003dr} in the $\Lambda$CDM and undulant
cosmologies.  We will present details elsewhere.

\begin{figure}
\includegraphics[width=8.5cm]{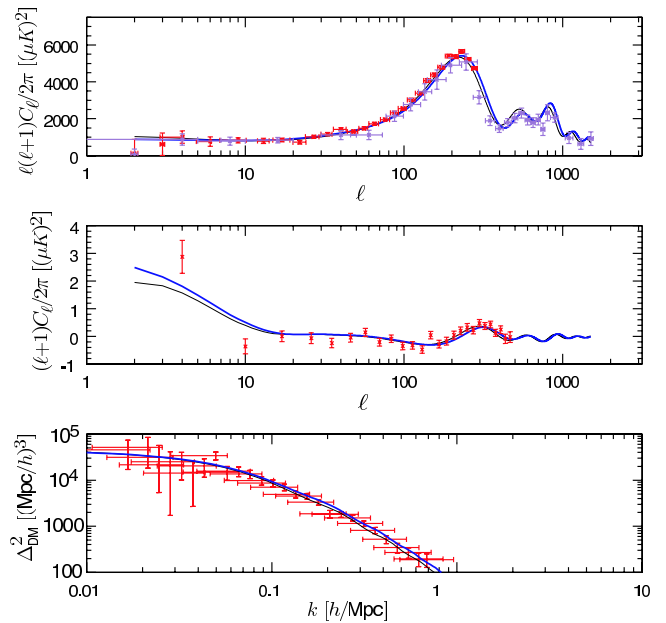}
\caption{Angular power spectrum (top panel) and $T$-$E$ cross-correlation (middle panel) 
versus the
multipole $\ell$, and matter power spectrum versus the wave 
number $k$ (bottom panel), for
the periodic equation of state (\ref{eq:oureos})  (blue line) and
for the $\Lambda$CDM model (black line).  The top panel shows
experimental data from the WMAP experiment
(red)~\cite{wmap} and  from the combination of all CMB data
(purple)~\cite{Wang:2002rt}.  The middle panel shows WMAP data. 
The data in the bottom panel are from an
independent analysis of the 2dF survey~\cite{Tegmark:2001jh}.}
\label{fig:cmbfast}
\end{figure}

We have modified the \textsc{cmbfast}~\cite{Seljak:1996is} code to compute 
the implications of the periodic equation of state for the 
anisotropies of the cosmic microwave background.  We show in 
Figure~\ref{fig:cmbfast} that the undulant universe describes the angular power spectrum,
temperature-polarization cross-correlation, and matter power spectrum with the same degree of
fidelity as the $\Lambda$CDM model.  

The undulant universe offers a new response to the cosmic 
coincidence problem: the current state of the Universe, with 
$\Omega_{m} \approx \Omega_{v}$ and $w_{v} \approx -1$, has happened 
before and will happen again. No fine tuning is required, in the sense 
that $0.5 \le \Omega_{v} \le 0.9$ with $w_{v} \le -0.7$ occurs with 
$\sim 10\%$ probability for $10^{-10} \le a \le 10^{+10}$. We have not 
given a physical picture for the stuff that makes up the vacuum 
energy, though the program for constructing a potential in which a 
scalar field has the required behavior is 
clear~\cite{Nakamura:1998mt,Gerke:2002sx}. 
On the observational front, 
it is of clear importance to seek evidence that the vacuum-energy 
equation of state varies with time~\cite{Alam:2003fg}, and to
find new constraints at different epochs.

Others have remarked~\cite{Krauss:1999br} that no finite set of
astronomical measurements made over a finite time will ever allow us to
determine the ultimate fate of our Universe, and have
quantified~\cite{Wang:2004nm} the limited reach of reliable
extrapolations.  The undulant universe explored here shows that the
range of possible destinies for the Universe, even in the near term, is
very broad indeed.  The universe need not necessarily evolve toward the
cataclysm of terminal inflation or recollapse, but might continue the
sedate drift of a big slink.

\begin{acknowledgments}
Fermilab is operated by Universities Research Association Inc.\ under
Contract No.\ DE-AC02-76CH03000 with the U.S.\ Department of Energy.
G.B. and O.M. 
acknowledge the stimulating environment of the Aspen Center for Physics.
It is a pleasure to thank Sean Carroll, Scott Dodelson, Joe Lykken, Gary Steigman,
Tim Tait,  Jochen Weller,  and Matias Zaldarriaga for enlightening conversations.
\end{acknowledgments}


\begin{thebibliography}{99}
    
    \bibitem{Riess:1998cb}
    A.~G.~Riess {\it et al.}  [Supernova Search Team],
    Astron.\ J.\  {\bf 116}, 1009 (1998)
    [arXiv:astro-ph/9805201].

    \bibitem{Perlmutter:1998np}
    S.~Perlmutter {\it et al.}  [Supernova Cosmology Project],
    Astrophys.\ J.\  {\bf 517}, 565 (1999)
    [arXiv:astro-ph/9812133].

    \bibitem{Bennett:2003bz}
    C.~L.~Bennett {\it et al.},
    Astrophys.\ J.\ Suppl.\  {\bf 148}, 1 (2003)
    [arXiv:astro-ph/0302207]. 
    A summary table of cosmological parameters is available at
    \url{http://lambda.gsfc.nasa.gov/product/map/wmap_parameters.cfm}.

    \bibitem{aussies}
    The Two-Degree Field Galaxy Redshift Survey obtained spectra for 
    $245\,591$ objects, mainly 
    galaxies; see
    \url{http://www.mso.anu.edu.au/2dFGRS/}.

    \bibitem{sloan}
    The Sloan Digital Sky Survey, \url{http://www.sdss.org}, measures
    distances to  $\mathcal{O}(10^{6})$ galaxies and quasars.

    \bibitem{Peebles:2002gy}
    P.~J.~E.~Peebles and B.~Ratra,
    Rev.\ Mod.\ Phys.\  {\bf 75}, 559 (2003)
    [arXiv:astro-ph/0207347].

    \bibitem{Freedman:2003ys}
    W.~L.~Freedman and M.~S.~Turner,
    Rev.\ Mod.\ Phys.\  {\bf 75}, 1433 (2003)
    [arXiv:astro-ph/0308418].

    \bibitem{Trodden:2004st}
    M.~Trodden and S.~M.~Carroll,
    ``TASI lectures: Introduction to Cosmology,''
    arXiv:astro-ph/0401547.

    \bibitem{kirshner}
    Robert P. Kirshner,
    \textit{The Extravagant Universe}
    (Princeton University Press, Princeton, 2002).

    \bibitem{Carroll:2001xs}
    S.~M.~Carroll,
    ``Dark Energy and the Preposterous Universe,''
    arXiv:astro-ph/0107571.
    See also S.~M.~Carroll, ``The Cosmological Constant,''
 Living Rev. Relativity \textbf{4,} 1Ê (2001), 
 \url{http://www.livingreviews.org/lrr-2001-1}.

 \bibitem{Bridle:2003yz}
 S.~L.~Bridle, O.~Lahav, J.~P.~Ostriker and P.~J.~Steinhardt,
 Science {\bf 299}, 1532 (2003)
 [arXiv:astro-ph/0303180].


 \bibitem{Krauss:1995yb}
 L.~M.~Krauss and M.~S.~Turner,
 Gen.\ Rel.\ Grav.\  {\bf 27}, 1137 (1995)
 [arXiv:astro-ph/9504003].

    \bibitem{Caldwell:1997ii}
    R.~R.~Caldwell, R.~Dave and P.~J.~Steinhardt,
    Phys.\ Rev.\ Lett.\  {\bf 80}, 1582 (1998)
    [arXiv:astro-ph/9708069].

    \bibitem{Susskind:2003kw}
    L.~Susskind,
    ``The anthropic landscape of string theory,''
    arXiv:hep-th/0302219.

    \bibitem{Steinhardt:2002ih}
    P.~J.~Steinhardt and N.~Turok,
    Science {\bf 296}, 1436 (2002).

    \bibitem{Feng:2004ff}
    B.~Feng, M.~Li, Y.~S.~Piao and X.~Zhang,
    ``Oscillating quintom and the recurrent universe,''
    arXiv:astro-ph/0407432.

    \bibitem{Eidelman:2004wy}
    S.~Eidelman {\it et al.}  [Particle Data Group],
    Phys.\ Lett.\ B {\bf 592}, 1 (2004).
    
       \bibitem{Dodelson:2001fq}
       S.~Dodelson, M.~Kaplinghat and E.~Stewart,
       Phys.\ Rev.\ Lett.\  {\bf 85}, 5276 (2000)
       [arXiv:astro-ph/0002360].

       \bibitem{Griest:2002cu}
       K.~Griest,
       Phys.\ Rev.\ D {\bf 66}, 123501 (2002)
       [arXiv:astro-ph/0202052].

    \bibitem{Cepa:2004bc}
    J.~Cepa,
    ``Constraints on the cosmic equation of state: Age conflict versus phantom
    energy. Age-redshift relations in an accelerated universe,''
    arXiv:astro-ph/0403616.

\bibitem{Bean:2002sm}
R.~Bean, S.~H.~Hansen and A.~Melchiorri,
Nucl.\ Phys.\ Proc.\ Suppl.\  {\bf 110}, 167 (2002)
[arXiv:astro-ph/0201127].

\bibitem{Riess:2004nr}
A.~G.~Riess {\it et al.}  [Supernova Search Team],
Astrophys.\ J.\  {\bf 607}, 665 (2004)
[arXiv:astro-ph/0402512].
For data analysis details see 
\url{http://www.astro.ucla.edu/~wright/}.

\bibitem{chandra}
The Chandra X-Ray Observatory, 
\url{http://chandra.nasa.gov}.

\bibitem{Allen:2004cd}
S.~W.~Allen, R.~W.~Schmidt, H.~Ebeling, A.~C.~Fabian and L.~van Speybroeck,
Mon.\ Not.\ Roy.\ Astron.\ Soc.\  {\bf 353}, 457 (2004)
[arXiv:astro-ph/0405340]; S.~W.~Allen, Private comunication.

\bibitem{Rapetti:2004aa}
D.~Rapetti, S.~W.~Allen and J.~Weller,
``Constraining Dark Energy with X-ray Galaxy Clusters, Supernovae and the
Cosmic Microwave Background,''
arXiv:astro-ph/0409574.

\bibitem{Linder:2003dr}
E.~V.~Linder and A.~Jenkins,
Mon.\ Not.\ Roy.\ Astron.\ Soc.\ {\bf 346}, 573 (2003)
[arXiv:astro-ph/0305286].

\bibitem{Seljak:1996is}
U.~Seljak and M.~Zaldarriaga,
Astrophys.\ J.\  {\bf 469}, 437 (1996)
[arXiv:astro-ph/9603033]; M.~Zaldarriaga, private communication.
The most recent version of the \textsc{cmbfast} code is at
\url{http://www.cmbfast.org/}.
    
    \bibitem{wmap}
    WMAP power spectrum data can be found at
    \url{http://lambda.gsfc.nasa.gov/product/map/}.

    \bibitem{Wang:2002rt}
    X.~Wang, M.~Tegmark, B.~Jain and M.~Zaldarriaga,
    Phys.\ Rev.\ D {\bf 68}, 123001 (2003)
    [arXiv:astro-ph/0212417].

    \bibitem{Tegmark:2001jh}
    M.~Tegmark, A.~J.~S.~Hamilton and Y.~Xu,
    Mon.\ Not.\ Roy.\ Astron.\ Soc.\  {\bf 335}, 887 (2002)
    [arXiv:astro-ph/0111575].



    \bibitem{Nakamura:1998mt}
    T.~Nakamura and T.~Chiba,
    Mon.\ Not.\ Roy.\ Astron.\ Soc.\  {\bf 306}, 696 (1999)
    [arXiv:astro-ph/9810447].

    \bibitem{Gerke:2002sx}
    B.~F.~Gerke and G.~Efstathiou,
    Mon.\ Not.\ Roy.\ Astron.\ Soc.\  {\bf 335}, 33 (2002)
    [arXiv:astro-ph/0201336].


    \bibitem{Alam:2003fg}
    U.~Alam, V.~Sahni, T.~D.~Saini and A.~A.~Starobinsky,
    arXiv:astro-ph/0311364.

    \bibitem{Krauss:1999br}
    L.~M.~Krauss and M.~S.~Turner,
    Gen.\ Rel.\ Grav.\  {\bf 31}, 1453 (1999)
    [arXiv:astro-ph/9904020].

    \bibitem{Wang:2004nm}
    Y.~Wang, J.~M.~Kratochvil, A.~Linde and M.~Shmakova,
    ``Current Observational Constraints on Cosmic Doomsday,''
    arXiv:astro-ph/0409264.

    
\end{thebibliography}

\end{document}